\documentclass[preprint,a4paper,3p,10pt,times,twocolumn]{elsarticle}

\usepackage{txfonts}
\usepackage{amsmath} % AMS Math Package
\usepackage{stmaryrd}
\usepackage{amssymb}	% Math symbols such as \mathbb
\usepackage{multirow}  % Allows for multiple rows
\usepackage{cases}
\usepackage[none]{hyphenat}

\usepackage{graphicx}% Include figure files
\usepackage{dcolumn}% Align table columns on decimal point
\usepackage{bm}% bold math
\usepackage{braket} % braket notation
\usepackage{color}
\usepackage{subfigure}
\def\beq{\begin{equation}}
\def\eeq{\end{equation}}
\def\beqa{\begin{eqnarray}}
\def\eeqa{\end{eqnarray}}
\def\bal#1\eal{\begin{flalign}#1\end{flalign}}
\def\bfig{\begin{figure}}
\def\efig{\end{figure}}

\newcommand{\eq}[1]{Eq.(#1)}

\newcommand{\fig}[1]{Fig.~#1}

\newcommand{\secc}[1]{Sec. #1}

\newcommand{\w}{\omega}

\newcommand{\pau}[1]{ {\sigma}_#1}
\newcommand{\pauj}[2]{ {{\sigma}^#2_#1}}

\newcommand{\ve}[1]{ \mathbf{\hat{r}}}

\newcommand{\abs}[1]{\left| #1 \right|} % for absolute value
\newcommand{\ketc}[1]{\left| #1 \right>} % for Dirac bras
\begin{document}
\begin{frontmatter}
%%\title{Transition from Jaynes-Cummings model to Rabi model with intermediate rotating wave approximation}
%\title{Transition from strong to ultrastrong coupling with intermediate RWA}
\title{Bridging the gap between the Jaynes-Cummings and Rabi models using an intermediate rotating wave approximation
}


%\author{Yimin~Wang$^{1,2}$ and Jing Yan Haw$^{2,3}$}
\author{Yimin~Wang$^{1,2}$}
%\ead{vivhappyrom@gmail.com}
\author{Jing Yan Haw$^{2,3}$}
%\ead{hawjingyan@gmail.com}

\address{$^1$College of Communications Engineering, PLA University of Science and Technology, Nanjing 210007, China\\
$^2$Centre for Quantum Technologies, National University of Singapore, Singapore\\
$^3$Centre for Quantum Computation and Communication Technology, Department of Quantum Science, The Australian National University, Canberra, ACT 0200, Australia}

%===============================================================================
\begin{abstract}
%%% We present a novel approach called the intermediate rotating wave approximation (IRWA), which employs a time-averaging method to address the co-rotating and counter-rotating terms in light-matter interaction separately. In contrast with the ordinary rotating wave approximation (RWA), this method allows one to trace their physical consequences individually, thus establishes the continuity between the Jaynes-Cummings model and the quantum Rabi model. We demonstrate IRWA in both near resonance and large detuning regime. The IRWA results not only agree well with both models in their respective coupling strengths, but also offers good explanation for the quantitative and qualitative differences between the models.

We present a novel approach called the intermediate rotating wave approximation (IRWA), which employs a time-averaging method to encapsulate the dynamics of light-matter interaction from strong to ultrastrong coupling regime. In contrast to the ordinary rotating wave approximation, this method addresses the co-rotating and counter-rotating terms separately to trace their physical consequences individually, and thus establishes the continuity between the Jaynes-Cummings model and the quantum Rabi model. We investigate IRWA in near resonance and large detuning cases. Our IRWA not only agrees well with both models in their respective coupling strengths, but also offers a good explanation for their differences.

\end{abstract}
%===============================================================================

\end{frontmatter}
%===============================================================================
%%%%%%%%%%%%%%%%%%%%%%%%%%%%%%%%%%%%%%%%%%%%%%%%%%%%%%%%%%%%%%%%%%%%%%%%%%%%%%%%%%%%%%%%%%%%%%%%%
% main text
\section{Introduction}
\label{sec_intro}
%Cavity quantum electrodynamics (cavity QED) studies the interaction between atoms and the quantized discrete electromagnetic modes inside a cavity. Such systems are of great interest and importance in quantum information processing and quantum optics \cite{mabuchi_cavity_2002}. The quantum electrodynamics of superconducting microwave circuits with Josephson junctions has been known as circuit QED by analogy to cavity QED in quantum optics, where the superconducting qubits play the role of artificial atoms and the microwave resonator plays the role of the cavity \cite{blais_cavity_2004, niemczyk_circuit_2010}.

The quantum Rabi model (QRM), which describes the interaction between a qubit and a quantized harmonic oscillator (bosonic mode) \cite{rabi_space_1937}, is written as
\bal
H_{Rabi}=\frac{1}{2}\hbar\omega_a \pau{z} +\hbar\omega_r a^\dag a +\hbar g \pau{x} (a+a^\dag),
\label{eq_h_rabi1}
\eal
where $a$ $(a^\dag)$ represents the bosonic annihilation (creation) operator of the electromagnetic field mode, $\omega_r$ is the corresponding frequency; $\omega_a$ is the transition frequency of the qubit, $\pau{i}$ $(i =x,y,z)$ are the corresponding Pauli operators; and $g$ is the dipole interaction strength. The QRM has been widely applied in modern physics, ranging from condensed-matter physics \cite{wagner_unitary_1986}, atomic physics \cite{cohen-tannoudji_atom-photon_1998} to quantum optics \cite{garrison_quantum_2008}, such as cavity QED \cite{raimond_manipulating_2001} and circuit QED \cite{blais_cavity_2004, niemczyk_circuit_2010} systems. Given its great importance, the QRM has been studied extensively using various methods \cite{irish_dynamics_2005, irish_generalized_2007}. Despite all those studies, the exact solution of QRM was only obtained by Braak recently \cite{braak_integrability_2011}. This analytical solution, however, is in the form of composite transcendental function defined in power series. The search for simpler analytical solution of generalized QRM with more physical insights thus continued \cite{zhong_analytical_2013, peng_solution_2014}. For example, the Bogoliubov-type transformation \cite{karpov_friedrichs_2000, flores-hidalgo_dressed-state_2002, passante_harmonic_2012}
is used to diagonalize the Hamiltonian to gain solutions and properties of the model \cite{chen_exact_2012, peng_exact_2013}.

The quantum Rabi model can be further simplified into the renowned Jaynes-Cummings model (JCM) \cite{jaynes_comparison_1963} by rotating wave approximation (RWA) provided that the coupling strength is sufficiently weak $(g \ll \min\{\omega_r,\omega_a\})$, and the detuning is small enough $(|\omega_a-\omega_r | \ll \omega_r+\omega_a)$. In the interaction picture, the rotating terms $\sigma_+ a$ and $\sigma_- a^\dag$ oscillate slowly with phase factor of $\text{exp}[\pm i (\omega_a-\omega_r) t]$, whereas the two ``counter-rotating'' terms $\sigma_+ a^\dag$ and $\sigma_- a$ oscillate rapidly with phase factor of $\text{exp}[\pm i ( \omega_r+\omega_a) t]$. Together with weak coupling condition, one can separate the time scales and discard the fast-oscillating terms \cite[p.354]{garrison_quantum_2008, zueco_qubit-oscillator_2009}, thus obtaining the Jaynes-Cummings Hamiltonian
\beq
\label{eq_h_jc1}
H_{JC}=\frac{1}{2}\hbar\omega_a \pau{z} +\hbar\omega_r a^\dag a +\hbar g (\pau{+}a + \pau{-}a^\dag ),
\eeq
which has simple analytical solutions.

Enhancement and tunability of light-matter interaction is crucial not only for fundamental studies of cavity/circuit QED but also for their applications in quantum information processing.
Three different coupling regimes can be defined based on the basic frequency scales of the system. In the weak coupling regime ($g \ll \{\gamma, \kappa\}$, $\kappa$ and $\gamma$ being the loss rate of the photon and the emitters' excitation), the discrete density of photonic states modifies the radiative lifetime of the quantum emitters (Purcell effect) \cite{purcell_spontaneous_1946}. Strong coupling regime is achieved when $\{\gamma, \kappa\} \ll g \ll \min\{\omega_r,\omega_a\}$ \cite[p.432]{meystre_elements_2007}, such that quantum emitters absorb and spontaneously re-emit a photon many times before dissipation becomes effective. This strong coupling regime has been investigated in various systems, ranging from atoms \cite{raimond_manipulating_2001}, through quantum dots (QD) \cite{khitrova_vacuum_2006} to Cooper-pair boxes \cite{blais_cavity_2004}. In these conventional QED experiments, the system is operating in either weak coupling regime or strong coupling regime. Therefore, the RWA, which leads to JCM from QRM, is very well justified, and the JCM captures a wealth of physical phenomena in conventional QED systems comprehensively.
With recent advances of new technologies, the ultrastrong coupling regime has become experimentally accessible in semiconductor \cite{gunter_sub-cycle_2009,geiser_ultrastrong_2012, scalari_ultrastrong_2012} and superconducting systems \cite{niemczyk_circuit_2010, forn-diaz_observation_2010}. In this so-called ultrastrong coupling regime \cite{casanova_deep_2010}, the coupling strength becomes comparable to the frequency of the resonator, $g/\omega_r \gtrsim 0.1$. Therefore, the routinely invoked RWA and the JCM break down, and the systems dynamics become governed by the QRM. This novel unexplored physics has opened up new research interest in applications of the QRM. Since then, considerable progress has been made and fascinating phenomenon have been predicted, such as photon blockade \cite{ridolfo_photon_2012}, nonclassical state generation \cite{ashhab_qubit-oscillator_2010}, breakdown of the standard master equation \cite{beaudoin_dissipation_2011}, and ultrafast two-qubit quantum gate operations \cite{romero_ultrafast_2012}.

One interesting observation is that the co-rotating terms and counter-rotating terms in the QRM affect the system in a different manner depending on the coupling regime. In light of this, one could gain better physical intuition of the continuity between the JCM and the QRM by treating the coupling strength of the co-rotating terms and counter-rotating terms separately \cite{shen_ground_2013, hioe_phase_1973}. In this paper, we seek to understand the emergence of the counter-rotating terms from the JCM to the QRM by resorting to the time-averaging method \cite[p. 353]{garrison_quantum_2008}, which also helps us to keep track of the time scale involved in the dynamics. With this method, we deploy a form of approximation, which we term as intermediate RWA (IRWA). The basic idea of this approximation is that, instead of going to the limit of either RWA or non-RWA, we use the time-averaged coupling strength in the interaction Hamiltonian. We present the general formalism and apply the IRWA into two specific situations: the near resonance case and the dispersive (large detuning) case.

The paper is organized as follows. In \secc{\ref{sec_tam_irwa}}, we first give a brief review of the time-averaging approach and introduce the IRWA.
In \secc{\ref{sec_res}}, the IRWA is used to study the energy levels of the system by perturbation theory in the near resonance case with increasing coupling strength.
In \secc{\ref{sec_dis}}, the dynamics of the system is investigated in the dispersive case with IRWA for both single- and multi-qubit case. We summarize our results in \secc{\ref{sec_sum}}.

\section{Time-averaging and intermediate RWA}
\label{sec_tam_irwa}
%Since the time scales in light-matter interaction are
%Why can we use time-averaging? work like
%Here we use the time-averaging method to separate the time (energy) scales involved in the system, and thus explore the the dynamics of those quantum systems.

\subsection{Time-averaging function}
The slow and fast time scales in a dynamical system can be separated explicitly by means of a temporal filtering operation. The time average of a function is defined by the convolution
\beq
\label{eq_run ave}
\overline{f(t)}=\int_{-\infty}^\infty dt' \varpi(t-t')f(t')=\int_{-\infty}^\infty dt' \varpi(t')f(t+t'),
\eeq
where the averaging function $\varpi(t)$ is positive, $\varpi(t)\geq 0 $, even, $\varpi(t)=\varpi(-t)$, and normalized, $\int_{-\infty}^\infty dt \,\varpi(t)=1 $ \cite[p.353]{garrison_quantum_2008}. The weighting function $\varpi(t)$ has a temporal width $ \tau = {\left[\int_{-\infty}^\infty \mathrm{d}t \varpi(t)\, t^2 \right]}^{1/2} < \infty$, which washes out oscillation with period smaller than $\tau$.

A simple example of such function is a Gaussian function
\beq
\varpi(t) = \frac{1}{ \tau \sqrt{2 \pi}} \; e^{- \frac{t^2}{2 \tau^2}}.
\label{eq_gaussian_weighting}
\eeq
It is more convenient to work in the domain of frequency, by using the Fourier transformed time-averaging function,
\beq
K(\omega)=\int^\infty_{-\infty} dt \,\varpi(t) e^{i\omega t},
\eeq
which is real and even, $K(-\omega)=K(\omega)=K^*(\omega)$, and has a finite width of $\omega_K\approx1/\tau$. The Fourier transform of the convolution \eq{\ref{eq_run ave}} is just the product of the individual Fourier transforms:
\beq
\overline{F(\omega)}=K(\omega)F(\omega).
\eeq
$K(\omega)$ is also called the cut-off function, which acts on $F(\omega)$ in such a way that $F(\omega)$ is essentially unchanged for small frequencies, $\omega \ll \omega_K$,  whereas frequencies larger than the width $\omega \gg \omega_K$ are strongly suppressed.

\subsection{Time-averaged Hamiltonian in intermediate RWA}
In order to apply the time-averaging function to the QRM, we impose the condition
\beq
g\ll \w_K
\label{eq:avecon}
\eeq
where the cut-off frequency $\w_K$ is chosen in such a way that the state interaction-picture state  $\ketc{\psi(t)}$ is essentially constant over the averaging interval, i.e.~$\ketc{\psi(t)}\approx\overline{\ketc{\psi(t)}}$. Upon time-averaging, the Schr\"{o}dinger equation of the QRM in the interaction picture can be written as \cite[p. 354]{garrison_quantum_2008}
%Since $\ketc{\psi(t)}$ only varies on the slow time scale, in interaction picture with respect to the free Hamiltonian $H_0=\hbar\omega_a \pau{z} /2 +\hbar\omega_r a^\dag a$, the time-averaging Schr\"{o}dinger equation can be written as \cite[p. 354]{garrison_quantum_2008}
\beq
i\hbar\frac{\partial}{\partial t}\ketc{\psi(t)}=\overline{H}_{int}\ketc{\psi(t)},
\label{eq_speq}
\eeq
where time-averaged Hamiltonian $\overline{H}_{int}(t)$ reads
\bal
\overline{H}_{int}(t)&=\overline{H}_{int,r}(t)+\overline{H}_{int,ar}(t), \\
\overline{H}_{int,r}(t)&=\hbar g_r\left( a\pau{+}e^{ i \Delta t}+a^\dag\pau{-}e^{-i \Delta t}\right), \\ \overline{H}_{int,ar}(t)&=\hbar g_{ar}\left( a\pau{-}e^{ -i \Sigma t}+a^\dag\pau{+}e^{i \Sigma t}\right).
\eal
Here, the time-averaged coupling strengths for co-rotating term $g_r$ and for counter-rotating term $g_{ar}$, are modified by the cut-off functions, such that $g_r=K(\pm\Delta)g$ and $g_{ar}=K(\pm\Sigma)g$, with $\Delta=\omega_a-\omega_r$ and $\Sigma=\omega_a+\omega_r$. This guarantees that the co-rotating terms $\sigma_+ a$, $\sigma_- a^\dag$ and the counter-rotating terms $\sigma_+ a^\dag$, $\sigma_- a$ contribute differently to the dynamics of the system, depending on the separation of the frequency scales. Notice that since the cut-off frequency $\w_K$ is coupling strength dependent (c.f.~\eq{\ref{eq:avecon}}), the cut-off function $K(\w)$ thus is a function of both $g$ and $\w$, i.e.~$K(\w,\w_K(g))$.

Going back to Schr\"{o}dinger picture, we then have the time-averaged quantum Rabi model as
\beq
\overline{H}_{Rabi}=H_0+\overline{H}_r+\overline{H}_{ar} ,
\label{eq_h_rabi2}
\eeq
where
\bal
\overline{H}_r&=\hbar g_r X_+ , &\text{with} && X_\pm &=a\pau{+}\pm a^\dag\pau{-};
\label{eq_hr_2} \\
\overline{H}_{ar}&=\hbar g_{ar} Y_+ , &\text{with} && Y_\pm&=a\pau{-}\pm a^\dag\pau{+}.
\label{eq_har_2}
\eal
The condition of $|\Delta| \leq \Sigma$ is generally satisfied in cavity/circuit QED systems, but this does not give a justification for us to neglect the contribution of the counter-rotating Hamiltonian $\overline{H}_{ar}$, and the QRM is still needed to describe the system. In the case of RWA, the sufficiently weak coupling condition, $g \ll \min\{\omega_r, \omega_a\} $, allows us to separate the frequency scales by
\beq
\label{eq_fresplit}
g \ll \omega_K \ll \min\{\omega_r, \omega_a\} \leq \Sigma.
\eeq
Therefore, when the coupling strength is weak compared to the free energy of the system, counter-rotating contribution is negligible since the cut-off function $K(\Sigma)$ is vanishingly small.
%Therefore, the contribution of the counter-rotating Hamiltonian is negligible in RWA due to the fact that its cut-off function $K(\Sigma)\approx 0$ is vanishing.
\begin{figure}[t]
\centering
\includegraphics[width=0.3\textwidth]{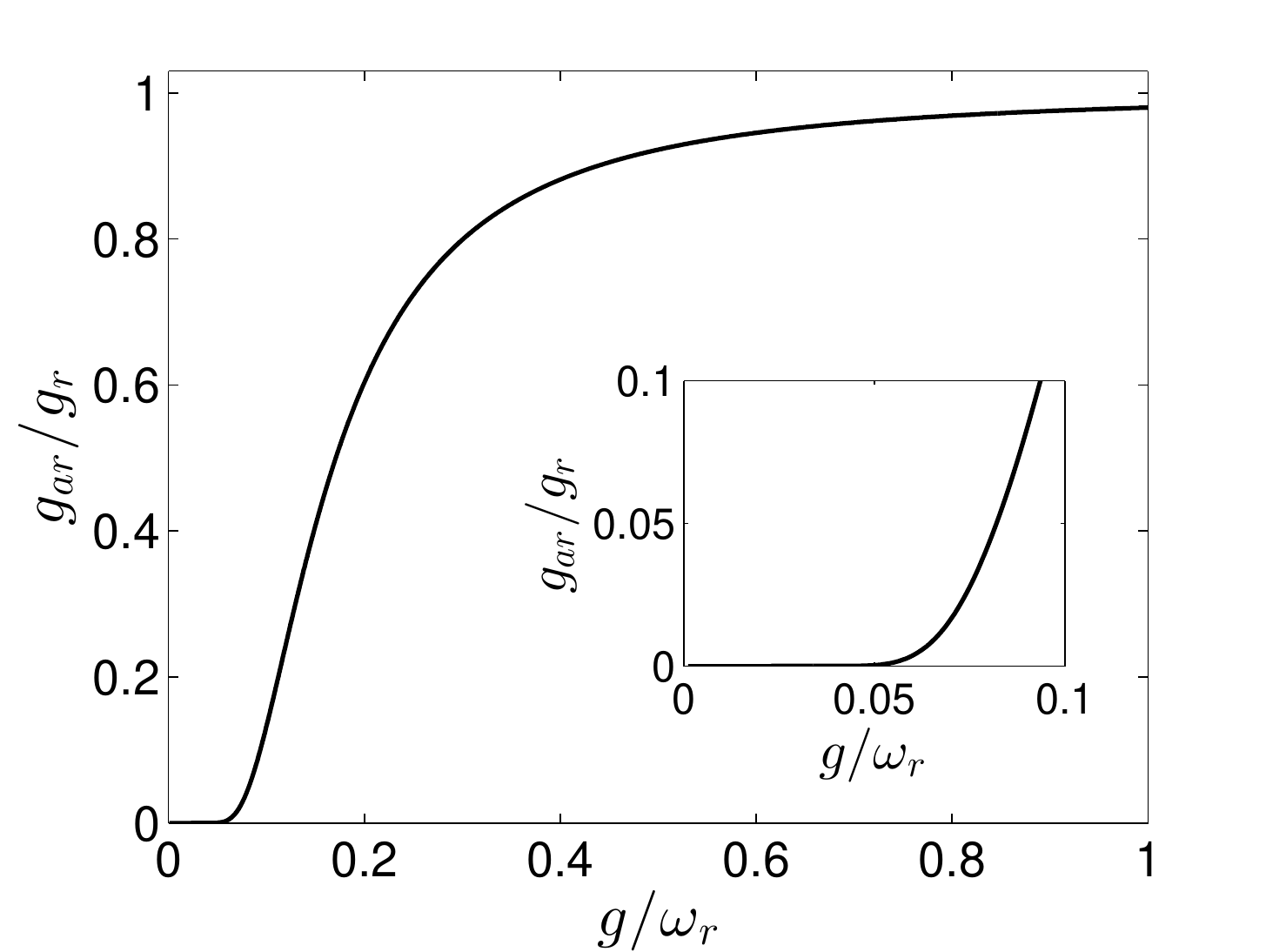}
\caption{The ratio of the time-averaged coupling strengths $g_{ar}$ for counter-rotating terms to $g_r$ for co-rotating terms as a function of coupling strength $g/\omega_r$, with Gaussian weighting function of width $\omega_K=10g$ for small detuning, $\Delta=0.01 \omega_r$. As the coupling strength is further increased, even though the near resonance condition is satisfied, the weak coupling condition is no longer respected and hence, RWA is not applicable any longer. The inset is a zoom of the region of $g/\omega_r$ between 0 and 0.1.}
\label{fig_gr_gar}
\end{figure}
In \fig{\ref{fig_gr_gar}}, we show the ratio between the two time-averaged coupling strengths $g_{ar}$ and $g_r$ as a function of the normalized coupling strength $g/\omega_r$, with Gaussian weighting function of width $\omega_K=10g$ for small qubit-resonator detuning $\Delta=0.01 \omega_r$. We note that, with small coupling strength, i.e.~$g/\omega_r \lesssim 0.05$, the time-averaged coupling strength $g_{ar}$ for counter-rotating terms is negligible for small detuning, which agrees well with the RWA conditions. We then arrive at the time-averaged JCM in Schr\"{o}dinger picture,
\beq
\label{eq_h_jc2}
\overline{H}_{JC}=\frac{1}{2}\hbar\omega_a\pau{z}+\hbar\omega_r a^\dag a+\hbar g_r X_+.
\eeq
However, as the coupling strength is getting larger, the contribution of the counter-rotating terms is increasing and hence, needs to be included to describe the dynamics correctly.

\section{Near resonance case in intermediate RWA}
\label{sec_res}
In the near resonance case of $\abs{\Delta} \ll \text{min} \{\omega_a, \omega_r\} \ll \omega_a + \omega_r$, the time-averaged coupling strength for counter-rotating terms is much smaller than the time-averaged coupling strength for co-rotating terms, $g_{ar}\ll g_r$. Thus, we take the time-averaged counter-rotating Hamiltonian in \eq{\ref{eq_har_2}} as a perturbation to the time-averaged Jaynes-Cummings Hamiltonian in \eq{\ref{eq_h_jc2}} and apply the non-degenerate stationary perturbation theory to obtain the second-order modification for energy \cite[p.249]{griffiths_introduction_2004}. The results can be written as

\bal
E_{0,g}&=E_{0,g}^{(0)}+E_{0,g}^{(1)}+E_{0,g}^{(2)},\\
E_{n,\pm}&=E_{n,\pm}^{(0)}+E_{n,\pm}^{(1)}+E_{n,\pm}^{(2)}, \,\,\, (n = 1,2,3...),
\eal
where
\bal
E_{0,g}^{(0)}&= -\frac{1}{2}\hbar \omega_a,\\
E_{n,\pm}^{(0)}&=\left(n+\frac{1}{2}\right)\hbar \omega_r \pm \frac{1}{2} \hbar \sqrt{\Delta^2 + 4 g_r^2 (n+1)}, \\
E_{0,g}^{(1)}&=E_{n,\pm}^{(1)}=0,
\eal
and
\bal
E_{0,g}^{(2)}&=\abs{\hbar g_{ar} }^2 \left(\frac{\abs{C_1 }^2}{E_{0,g}^{(0)}-E_{1,+}^{(0)}}+\frac{\abs{S_1}^2}{E_{0,g}^{(0)}-E_{1,-}^{(0)}}\right), \\
E_{0,+}^{(2)}&=2\abs{\hbar{g_{ar} S_0}}^2 \left(\frac{\abs{C_2}^2}{E_{0,+}^{(0)}-E_{2,+}^{(0)}}+\frac{\abs{S_2}^2}{E_{0,+}^{(0)}-E_{2,-}^{(0)}}\right), \\
E_{0,-}^{(2)}&=2\abs{\hbar {g_{ar} C_0}}^2 \left(\frac{\abs{C_2}^2}{E_{0,-}^{(0)}-E_{2,+}^{(0)}}+\frac{\abs{S_2}^2}{E_{0,-}^{(0)}-E_{2,-}^{(0)}}\right),
\eal
\bal
&E_{1,+}^{(2)}=\frac{\abs{\hbar g_{ar} C_1 }^2}{E_{1,+}^{(0)}-E_{0,g}^{(0)}}\nonumber \\
&+3\abs{\hbar {g_{ar} S_1}}^2 \left(\frac{\abs{C_3}^2}{E_{1,+}^{(0)}-E_{3,+}^{(0)}}+\frac{\abs{S_3}^2}{E_{1,+}^{(0)}-E_{3,-}^{(0)}}\right), \\
&E_{1,-}^{(2)}=\frac{\abs{\hbar g_{ar} S_1}^2}{E_{1,-}^{(0)}-E_{0,g}^{(0)}}\nonumber \\
&+3\abs{{\hbar g_{ar} C_1}}^2 \left(\frac{\abs{C_3}^2}{E_{1,-}^{(0)}-E_{3,+}^{(0)}}+\frac{\abs{S_3}^2}{E_{1,-}^{(0)}-E_{3,-}^{(0)}}\right),
\eal
\bal
E_{n\geq2,+}^{(2)}=n \abs{\hbar g_{ar} C_n}^2 \left(\frac{\abs{S_{n-2}}^2}{E_{n,+}^{(0)}-E_{n-2,+}^{(0)}}+\frac{\abs{C_{n-2}}^2}{E_{n,+}^{(0)}-E_{n-2,-}^{(0)}}\right)\nonumber \\
+(n+2) \abs{\hbar g_{ar} S_n}^2 \left(\frac{\abs{C_{n+2}}^2}{E_{n,+}^{(0)}-E_{n+2,+}^{(0)}}+\frac{\abs{S_{n+2}}^2}{E_{n,+}^{(0)}-E_{n+2,-}^{(0)}}\right),\\
E_{n\geq2,-}^{(2)}=n \abs{\hbar g_{ar} S_n}^2 \left(\frac{\abs{S_{n-2}}^2}{E_{n,-}^{(0)}-E_{n-2,+}^{(0)}}+\frac{\abs{C_{n-2}}^2}{E_{n,-}^{(0)}-E_{n-2,-}^{(0)}}\right)\nonumber \\
+(n+2) \abs{\hbar g_{ar} C_n}^2 \left(\frac{\abs{C_{n+2}}^2}{E_{n,-}^{(0)}-E_{n+2,+}^{(0)}}+\frac{\abs{S_{n+2}}^2}{E_{n,-}^{(0)}-E_{n+2,-}^{(0)}}\right),
\eal
with $C_n = \cos \theta_n$, $S_n = \sin \theta_n$, and $\theta_n = \arctan (2 g_r \sqrt{n+1}/\Delta)$.

\begin{figure}[t]
\begin{minipage}{0.49\linewidth}
  \leftline{\includegraphics[width=1\textwidth]{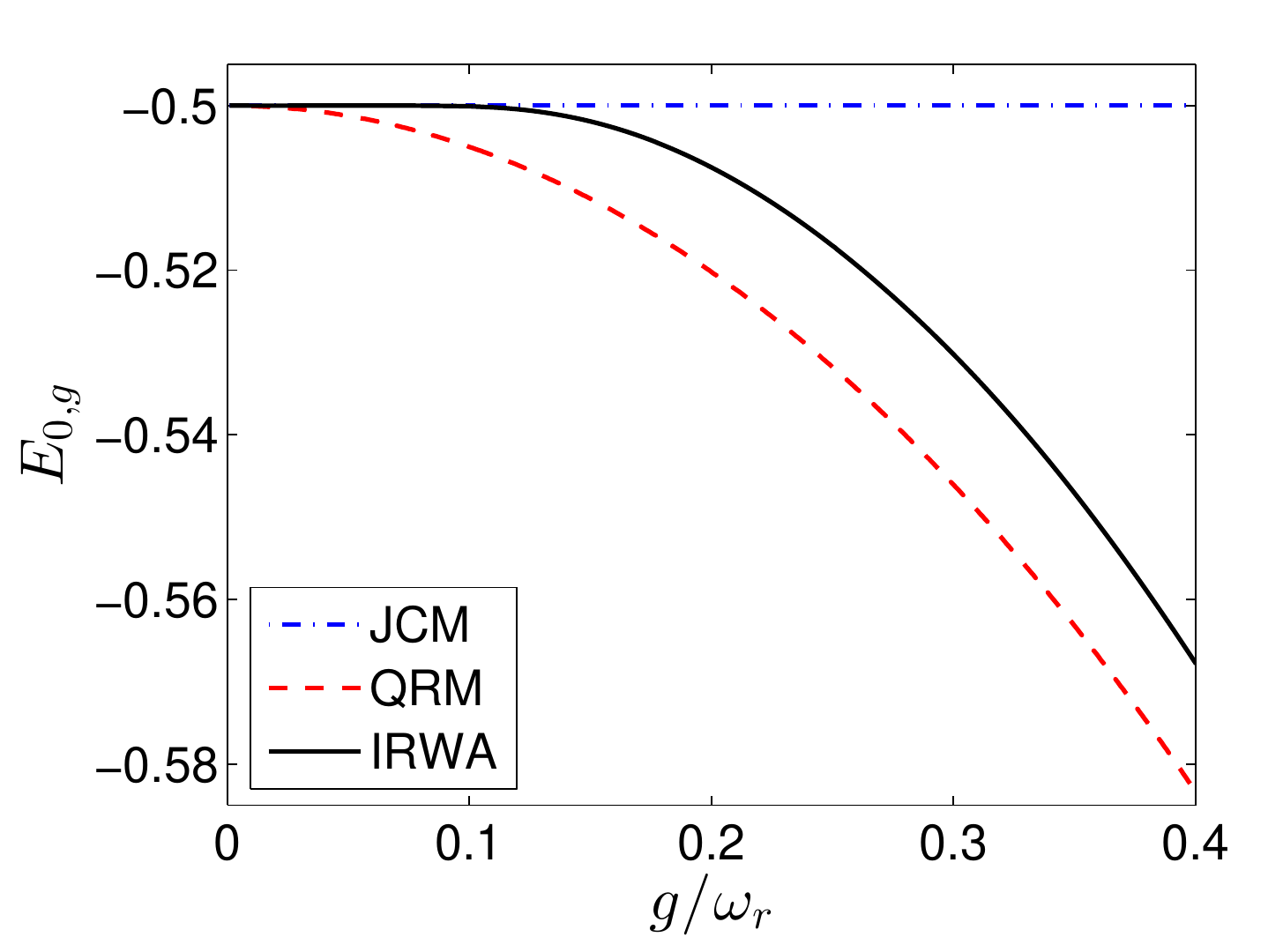}}
\vspace{0.0cm}
\end{minipage}
\hfill
\begin{minipage}{0.49\linewidth}
  \leftline{\includegraphics[width=1\textwidth]{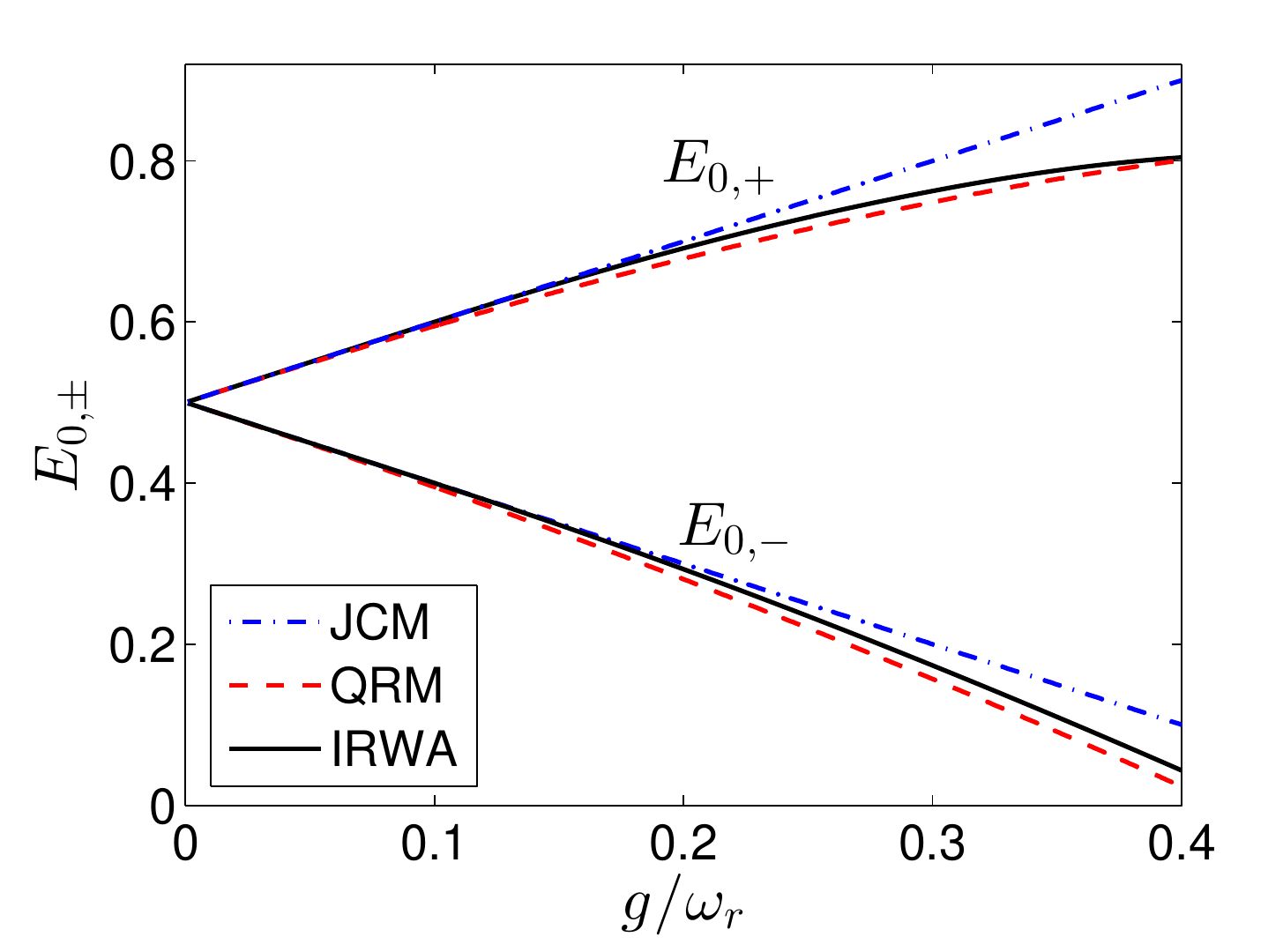}}
\vspace{0.0cm}
\end{minipage}
\begin{minipage}{0.49\linewidth}
  \leftline{\includegraphics[width=1\textwidth]{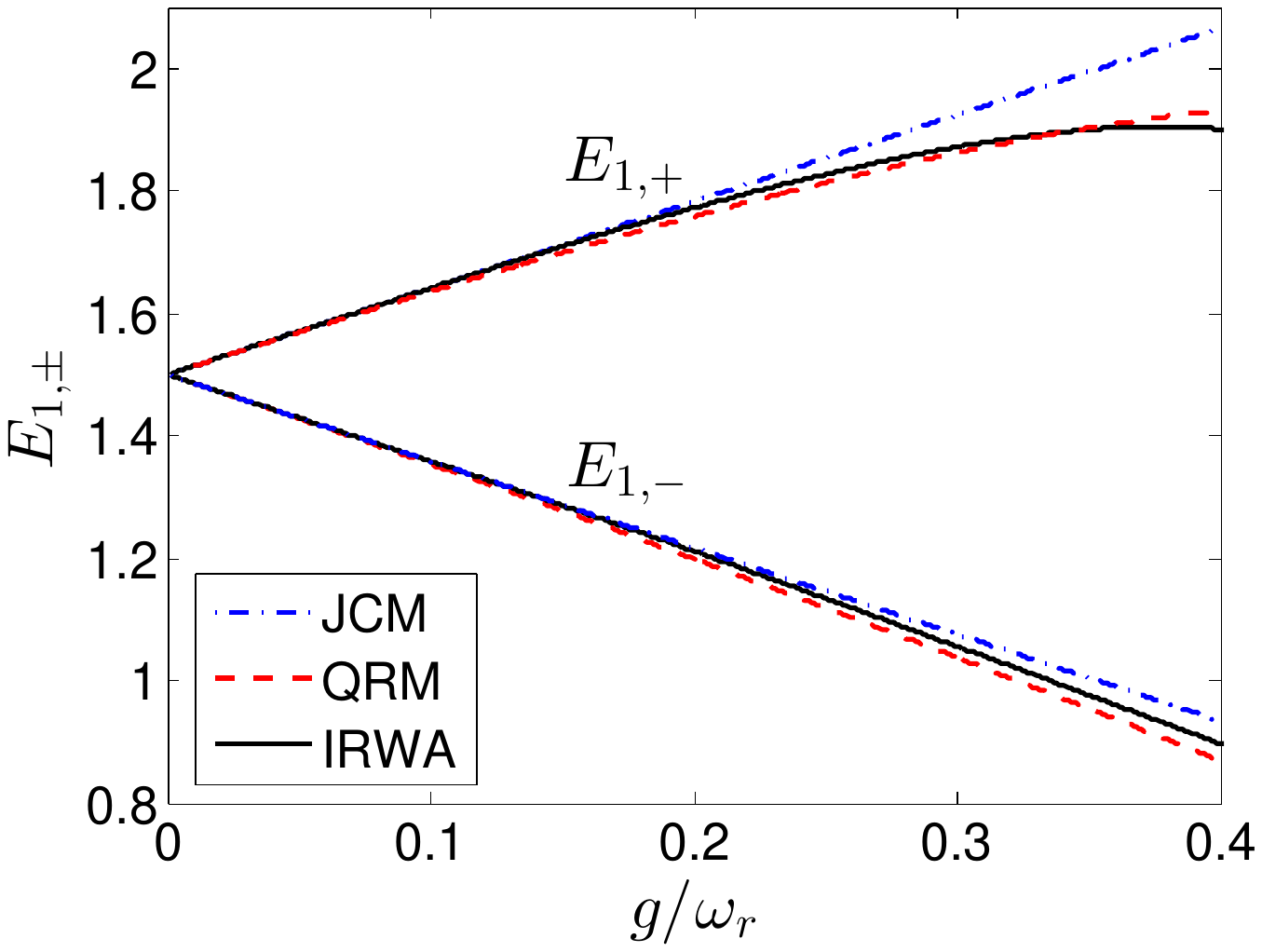}}
\vspace{0.0cm}
\end{minipage}
\hfill
\begin{minipage}{0.49\linewidth}
  \leftline{\includegraphics[width=1\textwidth]{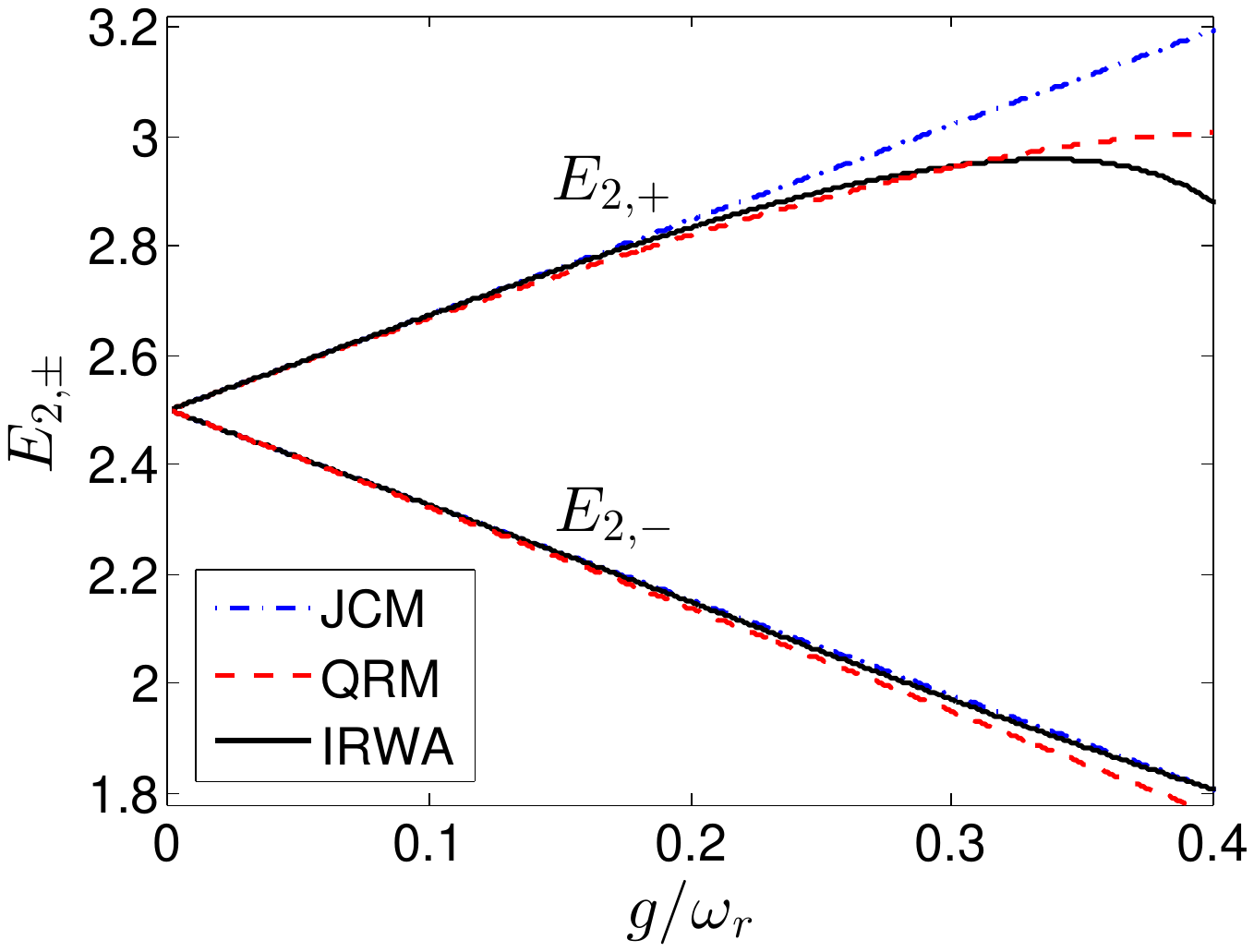}}
\vspace{0.0cm}
\end{minipage}
\caption{(Color online) The system spectrum as a function of the coupling strength $g/\omega_r$ with the time-averaged counter-rotating Hamiltonian of the IRWA in second perturbation theory (black solid lines) compared to the JCM (blue dashed-dotted lines) and the QRM (red dashed lines) for $\Delta=0$ and $\omega_K=10 g$.}
\label{fig_2ndper_energy}
\end{figure}

%The first-order modification for the wave functions is given in the Appendix.
The eigenenergies of the ground and low-lying excited states as a function of coupling strength $g/\omega_r$ are plotted in \fig{\ref{fig_2ndper_energy}}, which are
%for numerical results
obtained by our approach of IRWA (black solid lines), the JCM (blue dashed-dotted lines) and the QRM (red dashed lines). As shown in figure \fig{\ref{fig_2ndper_energy}}, the JCM curves deviate from the QRM curves for increasing coupling strength. The curves of the second-order perturbation theory in IRWA agree with the numerical results of the QRM for ultrastrong coupling regime of $g/\omega_r$ up to about $0.3$.

\section{Dispersive case in intermediate RWA}
\label{sec_dis}
%Although the analytical spectrum of Rabi Hamiltina has recently been found by Braak [26], it is defined in terms of the power series of a transcendental function. An approximate, but more simple form, can be found in the intermediate regime where g is small with respect to $\Sigma=\omega_a+\omega$, with the system still being in the ultrastrong coupling regime. This will be referred to as the Bloch-Siegert regime. This is done using the unitary transformation \cite{beaudoin_dissipation_2009}
In this section, we will study the dynamics of the Rabi Hamiltonian in the dispersive limit, where the qubit and the resonator are far detuned compared to the coupling strength $g\ll \abs{\Delta}$. The dispersive regime is of practical interests with applications in many cavity/circuit QED systems, such as quantum non-demolition measurement of the qubit \cite{guerlin_progressive_2007}, parity measurement of the two- or multi-qubit state \cite{lalumiere_tunable_2010}, and quantum gate operations \cite{majer_coupling_2007}. However, most of these applications were studied under the condition of strong coupling regime, where RWA still holds. In 2009, Zueco et al. generalized the studies of dispersive Hamiltonian to the ultrastrong coupling regime \cite{zueco_qubit-oscillator_2009}. Although the effect of counter-rotating terms is merely quantitative in single qubit case, the treatment beyond RWA gives rise to a qualitatively different effective model for multi-qubit scenario. Using the time-averaging functions in \secc{\ref{sec_tam_irwa}} to keep track of both the co-rotating and counter-rotating terms in the Hamiltonian, we can gain a better insight of their contributions and hence, the transition from the JCM to the QRM.

%developed from the JCM with RWA in strong coupling regime. Although the quantum Rabi model dynamics beyond RWA in dispersive case of ultrastrong regime has been studied \cite{zueco_qubit-oscillator_2009}, it is not clear how to map the transition between the two models with increasing coupling strength. So it is really worthy studying the dynamics of the system in dispersive regime using the time-averaging functions to keep track of both the ``co-rotating" and ``counter-rotating" term in the Hamiltonian.

\subsection{Dispersive Regime in One Qubit}
\label{sec_ged1q}
%The time-averaged quantum Rabi Hamiltonian is rewritten as
%\beq
%\overline{H}_{Rabi}=\frac{\hbar}{2}\omega_a\pau{z}+\hbar\omega a^\dag a+\hbar g_rX_++\hbar g_{ar}Y_+,
%\label{eq_h_rabi3}
%\eeq
%with
%\bal
%X_\pm&=a\pau{+}\pm a^\dag\pau{-}, \label{repx}\\
%Y_\pm&=a\pau{-}\pm a^\dag\pau{+}\label{repy}
%\eal
%\subsection{General Expression}
%We will now work out the dispersive regime in general, by incorporating the time-averaging functions introduced in Sec. \ref{tam} to keep track of both the co-rotating and counter-rotating term in the Hamiltonian. The full atom-field interaction, often denoted as \textit{Rabi Model} is
%\ba{
%H=H_0+H_I=\frac{\hbar}{2}\omega_a\pau{z}+\hbar\omega a^\dag a+\hbar g\left(a+a^\dag\right)\pau{x}}
%% where $H_0=\frac{\hbar}{2}\omega_a\pau{z}+\hbar\omega a^\dag a$, $H_I=\hbar g\left(a+a^\dag\right)\pau{x}$ and $\pau{x}=\pau{+}+\pau{-}$\footnote{A more general case is  $H_I=\hbar g\left(a+a^\dag\right)(\cos\theta\pau{z}-\sin\theta\pau{x})$. Nevertheless the $\pau{z}$ term can be discarded by applying polaron transformation \cite{reuther2009two}}.
%Once again we split the interaction Hamiltonian to co-rotating and counter-rotating parts
%\ba{H_I=\hbar g_rX_++\hbar g_{ar}Y_+}
%with
%
%and it is reminded that $ g_r=g K(\omega_a-\omega)$, $g_{ar}=g K(\omega_a+\omega)$, where $K(\omega)$ is the time-averaging function defined in Sec. \ref{tam}. Eventually the Hamiltonian takes the form

In the dispersive limit, where the coupling strength is much lesser than the qubit-resonator detuning,
\beq
g \ll \abs{\Delta},
\label{eq_dis}
\eeq
the time-averaged QRM in \eq{\ref{eq_h_rabi2}} can be transformed to
\bal
\label{eq_disp1}
H_{ir}^d=&U \overline{H}_{Rabi} U^\dag \nonumber \\
\approx & \frac{\hbar}{2}\omega_a \pau{z} +\hbar\omega_r a^\dag a+\frac{\hbar}{2}\left(\frac{g_r^2}{\Delta}+\frac{g_{ar}^2}{\Sigma}\right)\left[\pau{z}\left(2 a^\dag a+1\right)\right] \nonumber \\
&+\frac{\hbar}{2}g_r g_{ar}\left(\frac{1}{\Delta}+\frac{1}{\Sigma}\right)\left[\pau{z}\left( {a^\dag}^2+a^2 \right)\right],
\eal
up to second order in $\lambda=g_r/\Delta$ and $\Lambda=g_{ar}/\Sigma$ via the unitary transformation \cite{zueco_qubit-oscillator_2009,klimov_group-theoretical_2009},
\beq
U=\exp[\lambda X_-+\Lambda Y_-].
\eeq
$X_-$ and $Y_-$ are defined in \eq{\ref{eq_hr_2}} and \eq{\ref{eq_har_2}}. In strong coupling regime with RWA, we have the following inequalities
\bal
g \ll \abs{\Delta} \ll \omega_K \ll \Sigma,
\label{eq_dis_rwa}
\eal
which encapsulate the dispersive limit and the RWA conditions (near resonance and weak coupling limit) while respecting the time-averaging condition \eq{\ref{eq:avecon}}. When these inequalities are satisfied such that $g_{ar}\simeq 0$, $g_r=g$, the counter-rotating terms can be safely discarded. This gives rise to
\beq
H_r^d=\frac{\hbar}{2}\left(\omega_a+\frac{g^2}{\Delta}\right)\pau{z}+\hbar\left(\omega_r + \frac{g^{2}}{\Delta}\pau{z}\right)a^\dag a,
\label{eq_disp1_rwa}
\eeq
where the oscillator frequency is shifted as
\beq
\omega_r\rightarrow\omega_{r,r}=\omega_r \pm g^2/\Delta,
\label{eq_disp1_rwa_wcshift}
\eeq
depending on the state of the qubit. Similarly, the level separation of the qubit is shifted to
\beq
\omega_a\rightarrow\omega_{a,r}=\omega_a + \left( \frac{g^2}{\Delta} + 2 \frac{g^2}{\Delta} a^\dag a \right),
\label{eq_disp1_rwa_washift}
\eeq
depends on the number of photons in the resonator. The term $2 a^\dag a g^2/\Delta $, which is linear in the mean photon number $n=\langle a^\dag a\rangle$, is the ac-Stark shift \cite{blais_cavity_2004} and $g^2/\Delta$ is the Lamb shift \cite{zueco_qubit-oscillator_2009, wallraff_strong_2004}.
On the other hand, given the fact that the resulting Hamiltonian $H_r^d$ commutes with $\pau{z}$, i.e.\ $[H_r^d,\pau{z}]=0$, it allows quantum non-demolition measurement since the qubit's state will not be changed upon the evolution of the system. Hence, the state of the qubit can be inferred by probing the resonator frequency.

In ultrastrong coupling regime, where either or both of the RWA conditions are violated, we have the following inequalites instead,
\bal
g \ll \abs{\Delta} \leq \Sigma \ll \omega_K ,
\label{eq_dis_nrwa}
\eal
and all terms in \eq{\ref{eq_disp1}} will be retained, and the effective Hamiltonian then reads
\bal
\label{eq_disp1_nrwa}
H_{nr}^d=\frac{\hbar}{2}\omega_a\pau{z}+\hbar\left[\omega_r+ \frac{g^{2}}{2}\left(\frac{1}{\Delta}+\frac{1}{\Sigma}\right)\pau{z}\right]\left(a+a^\dag \right)^2.
\eal
This expression is analogous to the RWA dispersive Hamiltonian in \eq{\ref{eq_disp1_rwa}}, with an extra contribution of $\Sigma$ in the coupling term. However, this Hamiltonian is not diagonal in the eigenbasis of $H_0$ due to ${a^\dag}^2$ and $a^2$. Nevertheless, for $g/\omega_r<1$, we can reinterpret the result as the state-dependent shift of the resonator frequency's potential curvature $\omega_r^2$ \cite{zueco_qubit-oscillator_2009}.
%such that the oscillator frequency undergoes a shift according to $a^2(t)=a^2e^{-2i\omega_r t}$ and ${a^\dag}^2(t)={a^\dag}^ 2e^{2i\omega_r t}$. Thus we have
%\bal
%H_{nr}^d=\frac{\hbar}{2}\left[\omega_a+g^{2}\left(\frac{1}{\Delta}+\frac{1}{\Sigma}\right)\right]\pau{z}+\hbar\left[\omega_r+ g^{2}\left(\frac{1}{\Delta}+\frac{1}{\Sigma}\right)\pau{z}\right]a^\dag a.
%\eal
Hence, the dispersive Hamiltonian with non-RWA gives rise to a shift in the oscillator frequency of
\beq
\omega_r\rightarrow\omega_{r,nr}=\omega_r \pm g^2\left(\frac{1}{\Delta}+\frac{1}{\Sigma}\right),
\eeq
which implies that dispersive readout is also possible even in the ultrastrong coupling regime. Looking back at \eq{\ref{eq_disp1}}, we notice that both the time-averaged coupling strength $g_{ar}$ and $g_{ar}$ contribute to the two-photons terms ${a^\dag}^2$ and $a^2$ in the Hamiltonian. The time average coupling strength associated with counter-rotating terms also leads to an extra qubit dependent shift $g^{2}\pau{z}/\Delta$.

\begin{figure}[t]
\centering
\begin{minipage}{0.49\linewidth}
\leftline{\includegraphics[width=1\textwidth]{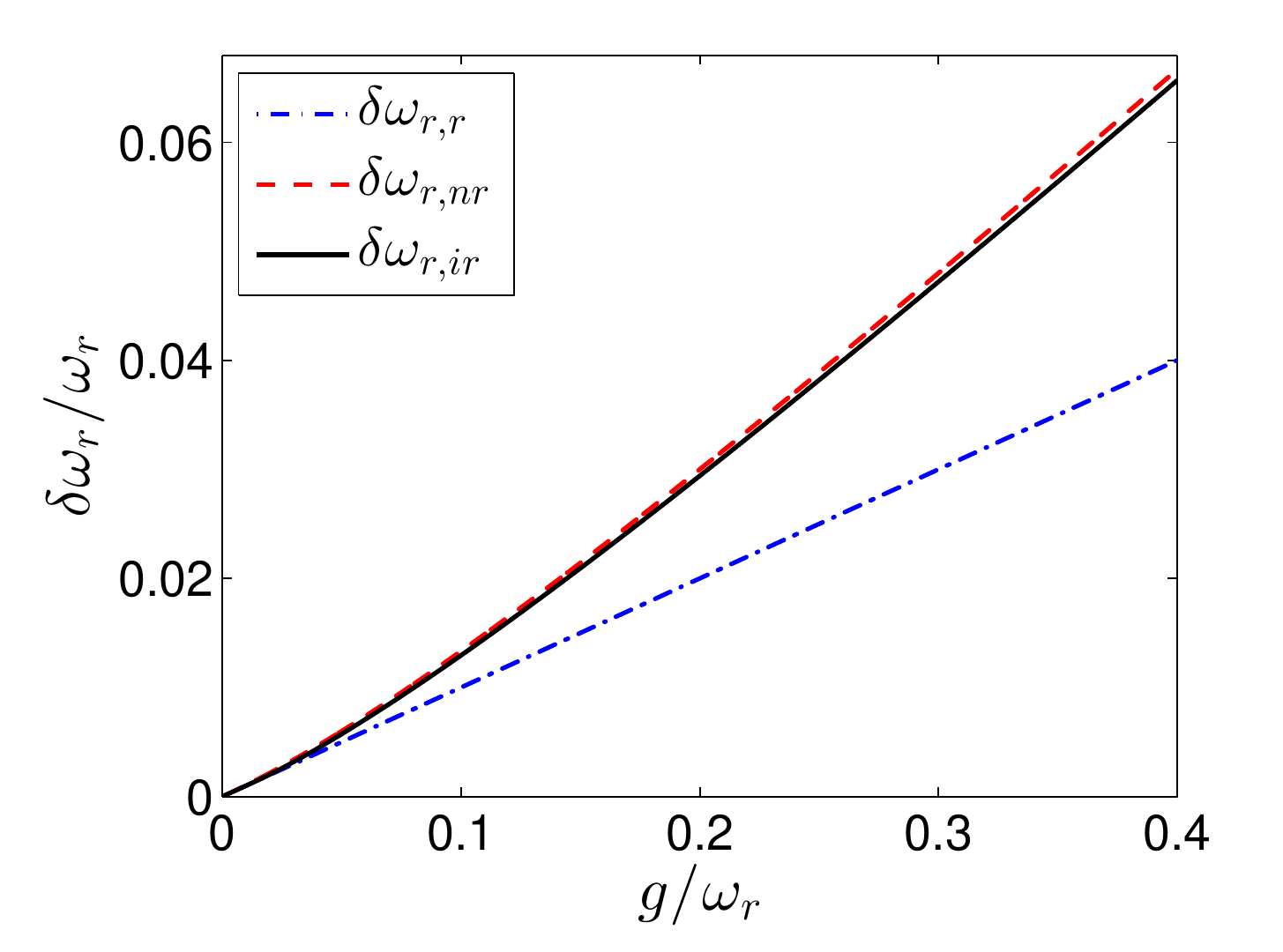}}
\vspace{0.0cm}
\end{minipage}
\hfill
\begin{minipage}{0.49\linewidth}
  \leftline{\includegraphics[width=1\textwidth]{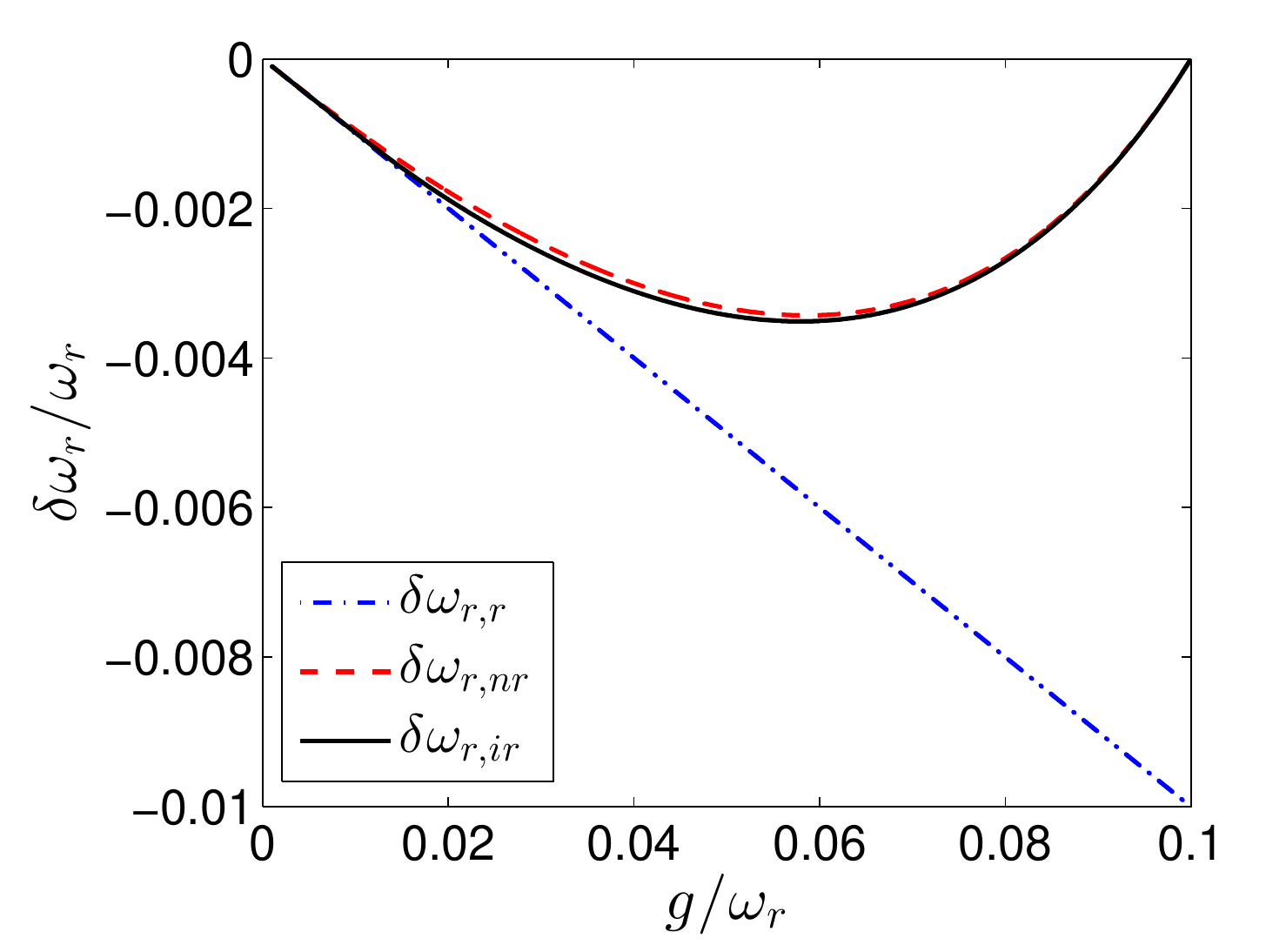}}
\vspace{0.0cm}
\end{minipage}
\caption{(Color online) Resonator frequency shift with single qubit in dispersive case as a function of coupling strength $g/\omega_r$ obtained with RWA (blue dashed-dotted lines), non-RWA (red dashed lines) and IRWA (black solid lines) for (a) positive detuning $\Delta>0$ ($\Delta=10g$) and (b) negative detuning $\Delta<0$ ($\Delta=-10g$) with Gaussian weighting function of width $\omega_K=10|\Delta|$. The breakdown of RWA is obvious, where it underestimates the dispersive shift for positive detuning and gives rise to a spurious shift in the absence of qubit ($\omega_a=0$) for negative detuning.}
\label{fig_irwa_11}
\end{figure}

Next, we study dynamics of the dispersive case from strong coupling regime to ultrastrong coupling regime use the time-averaged coupling strength in IRWA by numerical simulation. In \fig{\ref{fig_irwa_11}}, we show the frequency shift of the resonator as a function of normalized coupling strength $g/\omega_r$ for positive detuning $\Delta > 0$ and negative detuning $\Delta < 0$ in RWA (blue dashed-dotted lines), non-RWA (red dashed lines) and IRWA (black solid lines) with Gaussian weighting function.
It is clear that the RWA results have a totally different trend compared with the non-RWA results, especially for larger coupling strength. It underestimates the dispersive shift for positive detuning and predicts a shift even when the qubit's frequency $\omega_a$ tends to be zero for negative detuning ($\Delta \rightarrow \omega_r$ as g increases). This indicates the breakdown of RWA in predicting the dispersive resonator frequency shift in ultrastrong coupling regime. Meanwhile, our IRWA shows the manifestation of the counter-rotating terms as the coupling strength increases.
%has a good prediction of the the frequency shift as the non-RWA results that corresponds to the quantum Rabi model for both positive detuning in \fig{\ref{fig_irwa_11}} (a) and negative detuning in \fig{\ref{fig_irwa_11}} (b).

\begin{figure}[t]
\centering
\begin{minipage}{0.49\linewidth}
  \leftline{\includegraphics[width=1\textwidth]{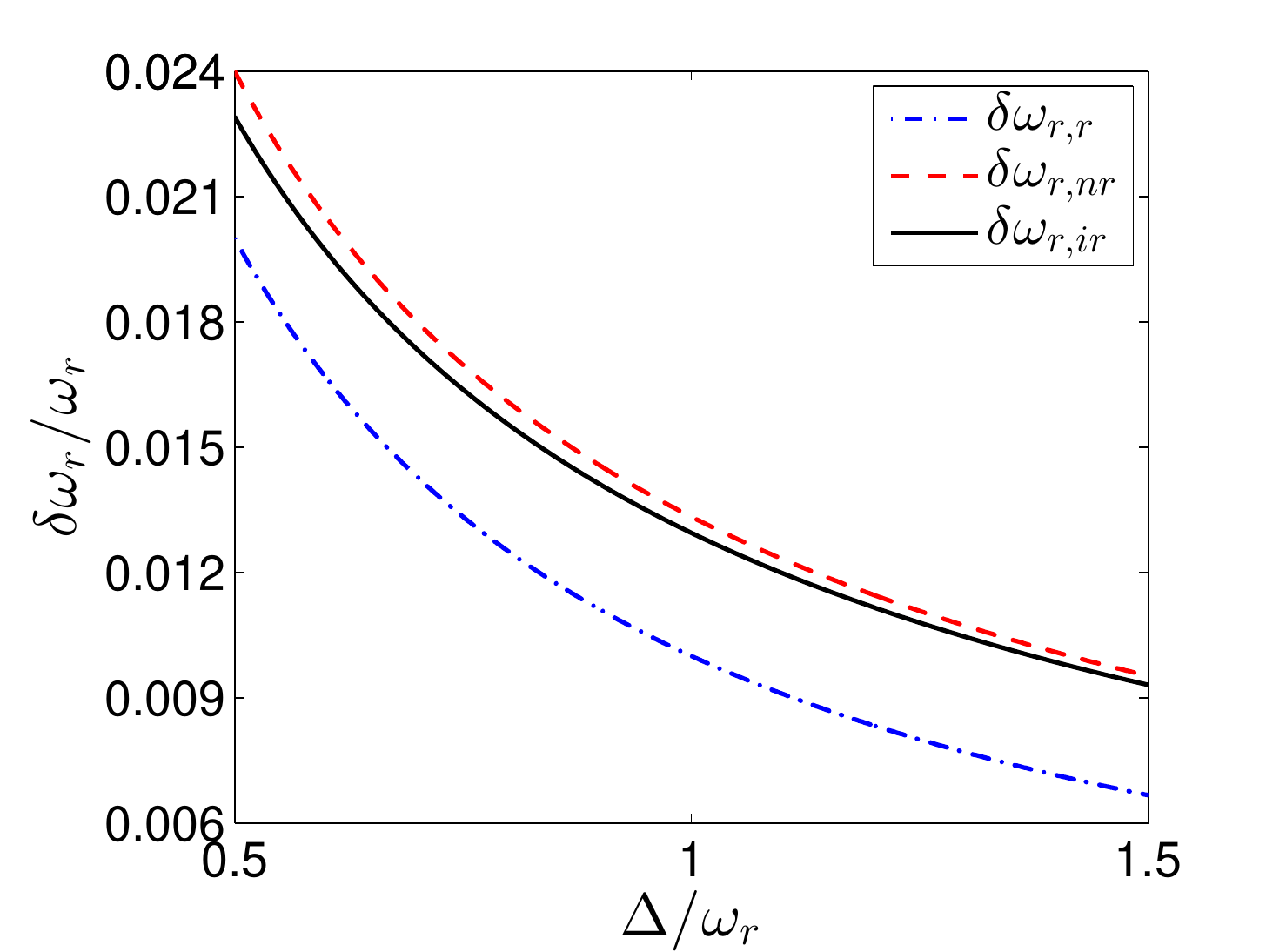}}
\vspace{0.0cm}
\end{minipage}
\hfill
\begin{minipage}{0.49\linewidth}
  \leftline{\includegraphics[width=1\textwidth]{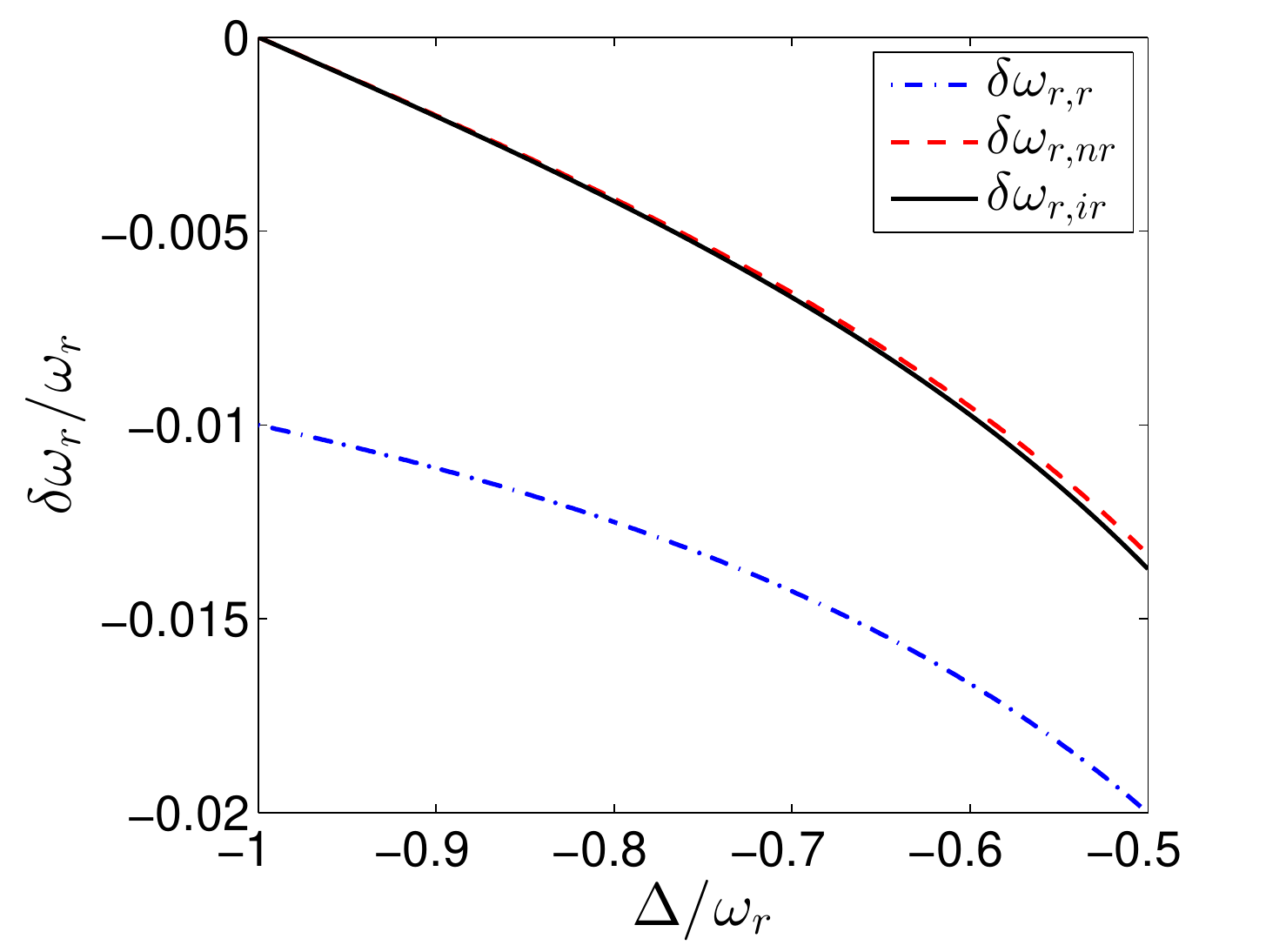}}
\vspace{0.0cm}
\end{minipage}
\caption{(Color online) Resonator frequency shift with single qubit in dispersive case as a function of detuning $\Delta/\omega_r$ obtained with RWA (blue dashed-dotted lines), non-RWA (red dashed lines) and IRWA (black solid lines) for (a) positive detuning $\Delta>0$ and (b) negative $\Delta<0$ with Gaussian weighting function of width $\omega_K=10|\Delta|$ and $g/\omega_r=0.1$.}
\label{fig_irwa_12}
\end{figure}

In \fig{\ref{fig_irwa_12}}, we show the frequency shift of the resonator as a function of detuning $\Delta/\omega_r$ in RWA (blue dashed-dotted lines), non-RWA (red dashed lines) and IRWA (black solid lines) with coupling strength of $g/\omega_r=0.1$ and Gaussian weighting function of width $\omega_K=10|\Delta|$. For this relatively large coupling strength, it is shown that the RWA results underestimate the resonator frequency shift for JCM, whereas the IRWA predictions agree quite well with the non-RWA results for QRM.
%It is easy to recognize that the RWA results for JCM underestimate the resonator frequency shift and the intermediate RWA predictions overlap with the non-RWA results for QRM quite well for this relatively large coupling strengths.

\subsection{Dispersive Regime with multi-qubit}
\label{sec_ged2q}
We now extend our discussion to multiple qubits coupled to a single mode resonator, where the time-averaged Hamiltonian takes the form \cite{blais_quantum-information_2007}
\beq
\label{eq_h2rabi2}
\overline{H}_{Rabi}^{nq}=\frac{\hbar}{2}\sum_{j}{\omega^j_a}{\pauj{z}{j}}+\hbar\omega_r a^\dag a+\hbar \sum_{j}\left(g^j_r X^j_++g^j_{ar} Y^j_+\right),
\eeq
with $X^j_\pm=a\pauj{+}{j}\pm a^\dag\pauj{-}{j}$ and $Y^j_\pm=a\pauj{-}{j}\pm a^\dag\pauj{+}{j}$. Applying the unitary transformation
\beq
U^{nq}=\exp\left(\sum_{j}\left(\lambda_j X^j_-+\Lambda_j Y^j_-\right)\right),
\eeq
and expanding the transformed Hamiltonian to the second order in $\lambda_j$ and $\Lambda_j$, we obtain the dispersive Hamiltonian
\bal
\label{eq_disp2_irwa}
H_{ir}^{2q,d}= &\hbar \omega_r a^\dag a + \frac{\hbar}{2} \sum_j \omega_a^j \sigma^j_z \nonumber \\
&+\frac{\hbar}{2} \sum_j \left(\frac{{(g_r^j)}^2}{\Delta_j}+\frac{{(g_{ar}^j)}^2}{\Sigma_j}\right)\left[\pauj{z}{j}\left(2 a^\dag a+1\right)\right] \nonumber \\
%&+\frac{\hbar}{2} \sum_j g_r^j g_{ar}^j \left(\frac{1}{\Delta_j}+\frac{1}{\Sigma_j}\right)\left[\pauj{z}{j}\left( {a^\dag}^2+a^2 \right)\right]  \nonumber \\
&+\frac{\hbar}{2}\sum_{j>k}g^j_r g^k_r\left(\frac{1}{\Delta_j}+\frac{1}{\Delta_k}\right)\left(\pauj{-}{j}\pauj{+}{k}+\pauj{+}{j}\pauj{-}{k}\right) \nonumber \\
&-\frac{\hbar}{2}\sum_{j>k}g^j_{ar}g^k_{ar}\left(\frac{1}{\Sigma_j}+\frac{1}{\Sigma_k}\right)\left(\pauj{-}{j}\pauj{+}{k}+\pauj{+}{j}\pauj{-}{k}\right) \nonumber \\
&+\frac{\hbar}{2}\sum_{j>k}g^j_rg^k_{ar}\left(\frac{1}{\Delta_j}-\frac{1}{\Sigma_k}\right)\left(\pauj{-}{j}\pauj{-}{k}+\pauj{+}{j}\pauj{+}{k}\right)\nonumber \\
&+\frac{\hbar}{2}\sum_{j>k}g^j_{ar}g^k_r\left(\frac{1}{\Delta_k}-\frac{1}{\Sigma_j}\right)\bigg]\left(\pauj{-}{j}\pauj{-}{k}+\pauj{+}{j}\pauj{+}{k}\right),
\eal
where the last four terms are the effective coupling between the qubits mediated by the resonator.
%\subsubsection{Case studies:RWA and NRWA}
%In the situation where the qubits are strongly detuned from each other, the effective coupling strength is essentially turned off since it does not conserve energy\cite{majer2007coupling,blais2007quantum}. This can be seen by going into the interaction picture in which each qubit is in the frame rotating at ${\omega^j_a}$. In this frame, when the qubits are strongly detuned, the interaction term, for instance $\pauj{+}{j}\pauj{-}{k}e^{i({\omega^j_a}-{\omega^k_0})t}$ is oscillating rapidly and averages out (See Sec. \ref{tam}), so the effective coupling is for all practical purposes turned off. Therefore, for maximum effective coupling, here we discuss the case where the qubits are degenerate with each other, ${\omega_a}_j={\omega_a}_k$ and hence $\Delta_j=\Delta_k=\Delta$ and $\Sigma_j=\Sigma_k=\Sigma$

%\subsubsection{rwa in 2 qubits}
To illustrate, we now take a two-qubit system as an example. In RWA, the counter-rotating terms are discarded because $g^{j,k}_{ar}\approx 0$. By setting all $g^j_r=g$ and $\omega_a^j=\omega_a$, we obtain
\bal
\label{eq_2rwa}
H_r^{2q,d}=&\frac{\hbar}{2}\sum_j\left( \omega_a^j+\frac{g^2}{\Delta}\right)\pauj{z}{j} +\hbar\sum_j\left(\omega_r+ \frac{g^{2}}{\Delta}\pauj{z}{j}\right)a^\dag a \nonumber \\
&+ \hbar \frac{g^2}{\Delta}\left(\pauj{-}{j}\pauj{+}{k}+\pauj{+}{j}\pauj{-}{k}\right),
\eal
where the interqubit interaction is of isotropic XY type, $\pauj{-}{j}\pauj{+}{k}+\pauj{+}{j}\pauj{-}{k}$. In a frame rotating at the qubit's frequency,  $H_r^{2q,d}$ generates the evolution
\bal
\label{eq_2rwau}
U_r^{2q,d}=&\text{exp}\left[ -i J_r t \left(a^\dag a +\frac{1}{2}\right)\left(\pauj{z}{j}+\pauj{z}{k}\right)\right] \nonumber \\
&\times\begin{pmatrix}
1 &0 &0& 0\\
0 &\cos J_r t &i\sin J_r t&0\\
0 &i\sin J_r t&\cos J_r t&0\\
0& 0&0&1
\end{pmatrix} \otimes I_r
\eal
with $I_r$ being the identity operator in resonator space and the effective coupling strength being
\beq
J_r=\frac{g^2}{\Delta}.
\eeq
This has been employed to generate qubit-qubit entanglement and quantum gate operations \cite{blais_cavity_2004,blais_quantum-information_2007}. For instance, by turning on the coupling for a period $t=\pi\Delta/4g^2$, we can generate a $\sqrt{i\textrm{SWAP}}$ gate which can be used to transform the state $\ketc{e_j,g_k}$ into an entangled state $1/\sqrt{2}\left(\ketc{e_j,g_k}+i\ketc{g_j,e_k}\right)$. Here, $\ketc{e_j}$ and $\ketc{g_k}$ are the excited state for $j$-th qubit and ground state for the $k$-th qubit, respectively.

In ultrastrong coupling regime without RWA, all the terms will be retained. By setting all $g^j_r=g^j_{ar}=g$ and $\omega_a^j = \omega_a$ to be equal, we obtain
\bal
\label{eq_2nrwa}
H_{nr}^{2q,d}&=\frac{\hbar}{2}\sum_j\left[\omega_a^j+g^{2}\left(\frac{1}{\Delta}+\frac{1}{\Sigma}\right)\right]\pauj{z}{j} \nonumber \\&+\hbar g^2\left(\frac{1}{\Delta}-\frac{1}{\Sigma}\right)\pauj{x}{j}\pauj{x}{k} \nonumber \\
&+\hbar\sum_j\left[\omega_r+g^{2}\left(\frac{1}{\Delta}+\frac{1}{\Sigma}\right)\pauj{z}{j}\right]a^\dag a,
\eal
where the interqubit interaction is of Ising type $\pauj{x}{j}\pauj{x}{k}$. The evolution operator reads
\bal
\label{eq_2nrwau}
&U_{nr}^{2q,d}=\text{exp}\left[ -i J_{nr,0} t \left(a^\dag a +\frac{1}{2}\right)\left(\pauj{z}{j}+\pauj{z}{k}\right)\right] \nonumber \\
&\times \begin{pmatrix}
\cos J_{nr,1} t &0 &0& i\sin J_{nr,1} t\\
0 &\cos J_{nr,1} t &i\sin J_{nr,1} t&0\\
0 &i\sin J_{nr,1} t&\cos J_{nr,1} t&0\\
i\sin J_{nr,1} t& 0&0&\cos J_{nr,1} t
\end{pmatrix}\otimes I_r,
\eal
in the frame rotating at the qubit's frequency with the effective coupling strength being
\bal
J_{nr,0}&=g^2\left(\frac{1}{\Delta}+\frac{1}{\Sigma}\right),\\
J_{nr,1}&=g^2\left(\frac{1}{\Delta}-\frac{1}{\Sigma}\right).
\eal
It is worth noting that the extension from \eq{\ref{eq_2rwa}} to \eq{\ref{eq_2nrwa}} is not just a renormalization of the parameters. The effective qubit-qubit interaction type is indeed different, which will be clearer when we compare the evolution operators for RWA and non-RWA (\eq{\ref{eq_2rwau}} and \eq{\ref{eq_2nrwau}}), where one is isotropic XY interaction while the other is Ising type interaction respectively. To understand this apparent sudden transition between RWA and non-RWA in dispersive regime for the multi-qubit, we invoke the time-averaged IRWA interpretation. From the effective Hamiltonian in \eq{\ref{eq_disp2_irwa}}, the evolution operator can be written as
\bal
\label{eq_2irwau}
&U_{ir}^{2q,d}=\text{exp}\left[ -i J_{ir,0} t \left(a^\dag a +\frac{1}{2}\right)\left(\pauj{z}{j}+\pauj{z}{k}\right)\right] \nonumber \\
&\times \begin{pmatrix}
\cos J_{ir,2} t &0 &0& i\sin J_{ir,2} t\\
0 &\cos J_{ir,1} t &i\sin J_{ir,1} t&0\\
0 &i\sin J_{ir,1} t&\cos J_{ir,1} t&0\\
i\sin J_{ir,2} t& 0&0&\cos J_{ir,2} t
\end{pmatrix}\otimes I_r,
\eal
where the effective coupling strengths being
\bal
J_{ir,0}&=\frac{{(g_r^j)}^2}{\Delta_j}+\frac{{(g_{ar}^j)}^2}{\Sigma_j},\\
J_{ir,1}&=g^j_rg^k_r\left(\frac{1}{\Delta_j}+\frac{1}{\Delta_k}\right)-g^j_{ar}g^k_{ar}\left(\frac{1}{\Sigma_j}+\frac{1}{\Sigma_k}\right),\\
J_{ir,2}&=g^j_rg^k_{ar}\left(\frac{1}{\Delta_j}-\frac{1}{\Sigma_k}\right)+g^j_{ar}g^k_r\left(\frac{1}{\Delta_k}-\frac{1}{\Sigma_j}\right).
\eal
%With the constraints under the consideration, here we show one of the examples of the IRWA plot, which is the case where detuning $\Delta$ is positive and we decrease $\omega$, keeping $\omega_a$ constant. In the sense of subspace which the terms span, $J_{ir,1}$ corresponds to RWA part and $J_{ir,2}$ corresponds to the non-RWA part.\\

\begin{figure}[t]
\centering
\begin{minipage}{0.49\linewidth}
  \leftline{\includegraphics[width=1\textwidth]{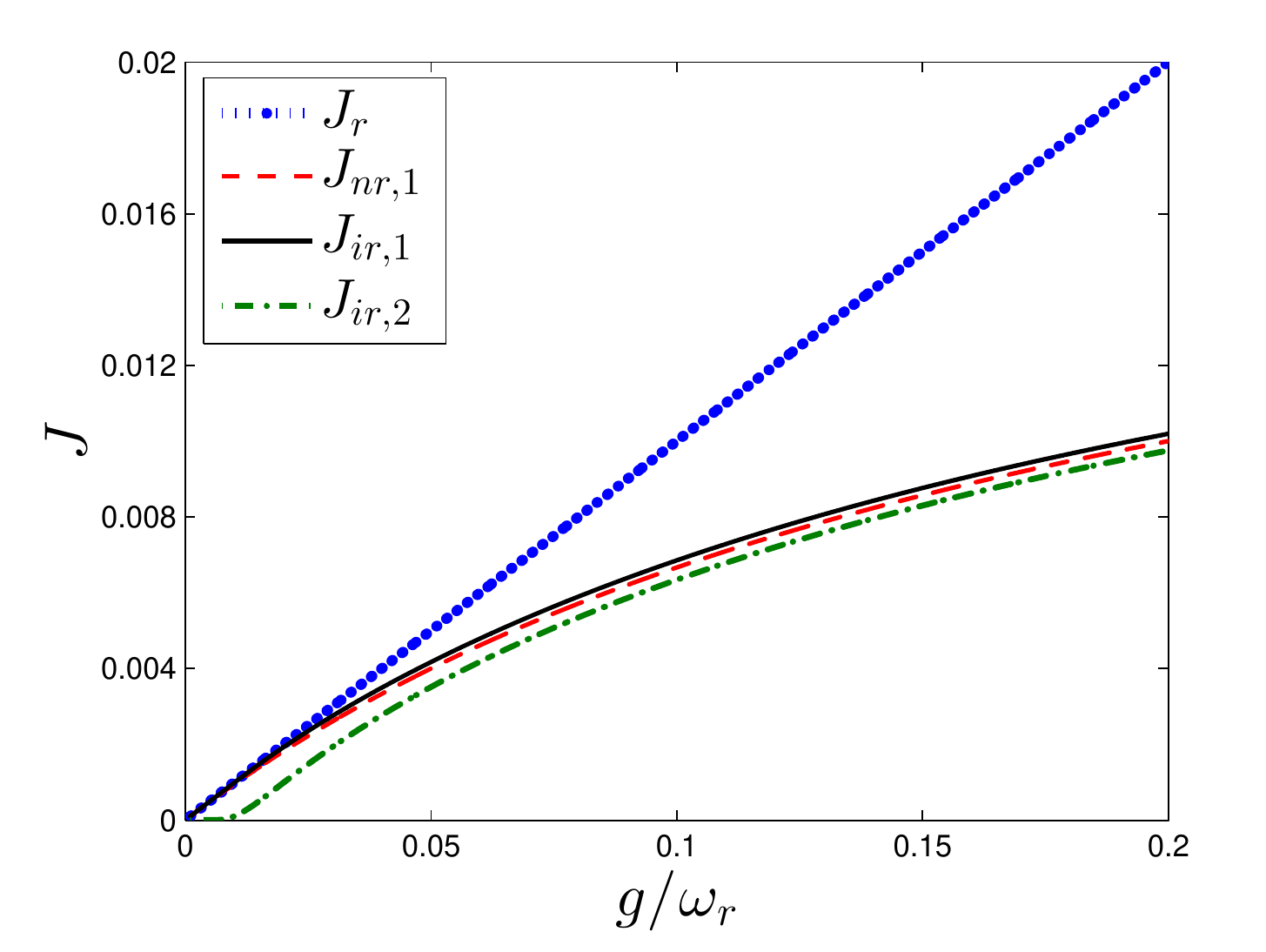}}
\vspace{0.0cm}
\end{minipage}
\hfill
\begin{minipage}{0.49\linewidth}
  \leftline{\includegraphics[width=1\textwidth]{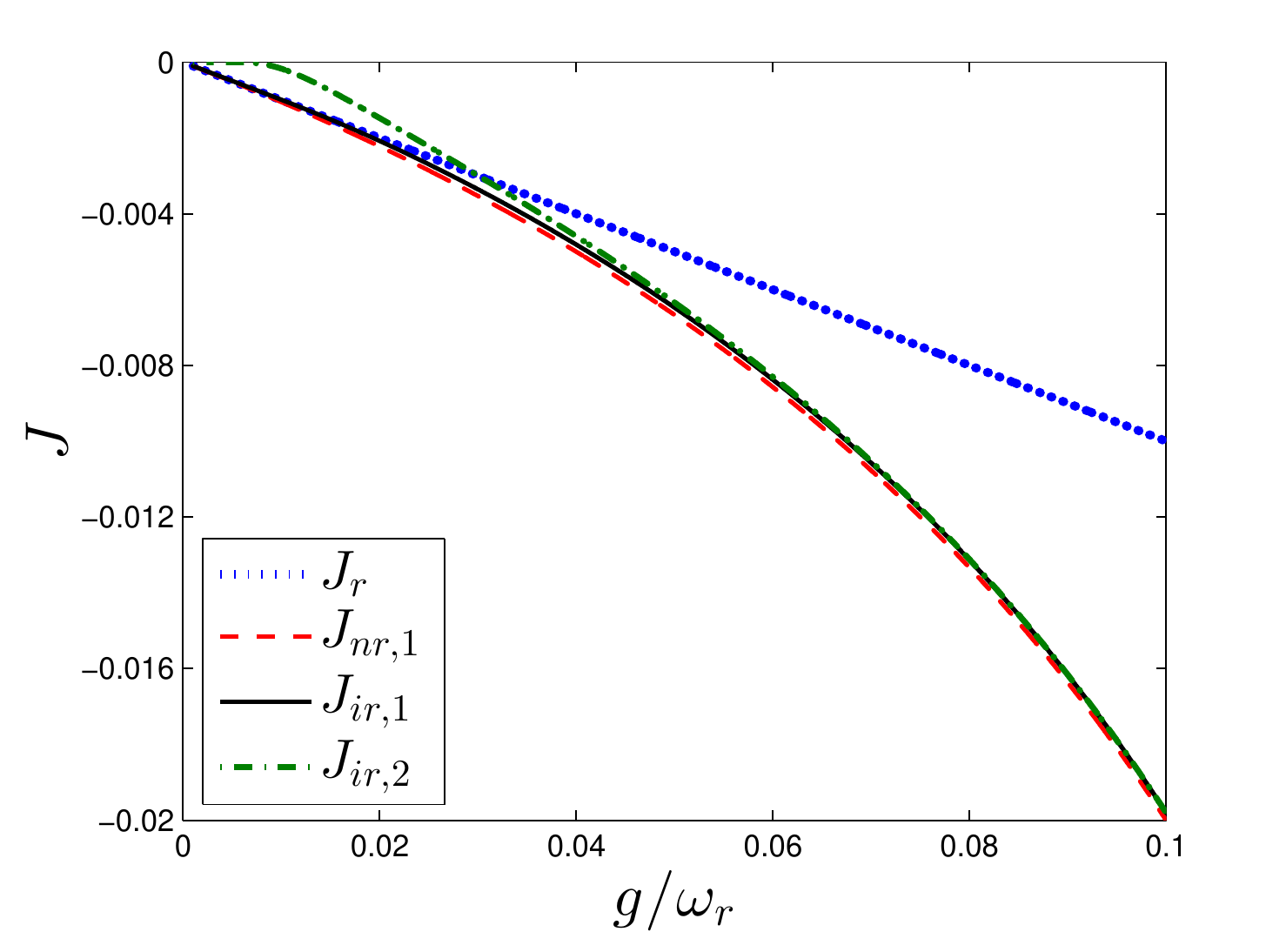}}
\vspace{0.0cm}
\end{minipage}
\caption{(Color online) The effective coupling strength in two-qubit dispersive case as a function of coupling strength $g/\omega_r$ obtained with RWA (blue dotted lines), non-RWA (red dashed lines) and IRWA (black solid lines and green dashed-dotted lines) for (a) positive detuning $\Delta>0$ ($\Delta=10g$) and (b) negative detuning $\Delta<0$ ($\Delta=-10g$) with Gaussian weighting function of width $\omega_K=10|\Delta|$.}
\label{fig_irwa_2}
\end{figure}

In \fig{\ref{fig_irwa_2}}, we show the transition of the difference for the qubit-qubit interaction type from RWA case to non-RWA using the IRWA with Gaussian weighting function of width $\omega_K=10|\Delta|$. As we can see, when the coupling strength is very small, $g/\omega_r\ll 0.1$, the IRWA curves are closer to the RWA curves for $J_{ir,1}$, whereas they are essentially zero for $J_{ir,2}$.
%the IRWA curves for $J_{ir,1}$ are closer to the RWA curves while for $J_{ir,2}$, it is essentially zero.
When the coupling strength is increased, the counter-rotating term coupling strength $g_{ar}$ starts to become significant, and hence, leads to the correction of $J_{ir,1}$ and manifestation of $J_{ir,2}$. As the coupling strength further increases, our IRWA curves start to deviate from the RWA curves and agree better with the non-RWA curves, showing the transitions from RWA to non-RWA. Eventually as the coupling strength reaches ultrastrong coupling regime, we regain the non-RWA results as in Eq. \eqref{eq_2nrwau}.

\section{Conclusion}
\label{sec_sum}
In this paper, we introduced the IRWA that is based on the time-averaging method for better understanding of the roles of the ``counter-rotating" terms in the QRM and the transition between strong coupling and ultrastrong coupling regimes. The eigenenergies of the system were studied by combining the perturbation theory and IRWA for near resonance case. The results agreed well with the JCM predictions for small coupling strength, i.e. $g/\omega_r$ up to 0.1 and with the QRM results for larger coupling strength, i.e. $g/\omega_r$ up to 0.3. We also showed that in dispersive regime, our IRWA predication gave a good explanation of the qubit-dependent frequency shifts in the single qubit scenario. This approach revealed the emergence of counter-rotating terms in the interqubit coupling, which leads to both quantitative and qualitative differences in the interaction strength and interaction type. Compared with other approaches \cite{zhong_analytical_2013, peng_solution_2014, chen_exact_2012, peng_exact_2013}, our IRWA method allows us to gain the physical consequences of the co-rotating and counter-rotating coupling terms individually by tracing those terms separately. As a remark, there are several aspects that still can be explored with the idea of IRWA. For instance, by relating the measurement interval to the width of the time-averaging function $\omega_K$ in our analysis, we can extend the result in \cite{lizuain_zeno_2010} to observe the transition from quantum Zeno effect to quantum anti-Zeno effect \cite{zheng_quantum_2008, ai_quantum_2010}. Our IRWA approach could also be applied to the studies of applicability of RWA in various phenomena, such as
%Quantum Zeno/Anti-Zeno effect \cite{zheng_quantum_2008,ai_quantum_2010} and
Berry phase in quantum systems \cite{larson_absence_2012, deng_berry_2013}, asymmetric couplings \cite{peng_solution_2014}, and generalized multi-qubit quantum Rabi model \cite{shen_ground_2013}. The IRWA might be useful as well in studying the dynamics of multiple coupling regimes in a single system, for example, by having one qubit coupled strongly in RWA regime and the other one operated ultrastrongly beyond RWA regime.

\section{Acknowledgements}
We acknowledge Valerio Scarani for his proposal of this research topic and valuable comments. We thank Tammy Chin for suggestions and feedback on the manuscript.
This work was supported by Natural Science Foundation of Jiangsu Province (No. BK20140072) and National Natural Science Foundation of China (No. 11404407). J. Y. Haw would like to acknowledge the support of the Australian Research Council Centre of Excellence for Quantum Computation and Communication Technology (project number CE110001027).

%
%\bibliographystyle{elsarticle-num-names}
%\bibliography{qo_ref_2014}

\end{document}